\documentclass[aps,prb,showpacs,twocolumn,groupedaddress]{revtex4}
\usepackage[latin1]{inputenc}
\usepackage{graphicx}
\usepackage{color}

\begin{document}


\title{Neutron scattering study of PbMg$_{1/3}$Ta$_{2/3}$O$_3$ and 
BaMg$_{1/3}$Ta$_{2/3}$O$_3$ complex perovskites}



\author{S.N. Gvasaliya}
\altaffiliation{On leave from Ioffe Physical Technical Institute, 
26 Politekhnicheskaya, 194021, St. Petersburg, Russia}
\affiliation{Laboratory for Neutron Scattering ETHZ \& Paul-Scherrer
Institut CH-5232 Villigen PSI Switzerland}

\author{B. Roessli}
\affiliation{Laboratory for Neutron Scattering ETHZ \& Paul-Scherrer
Institut CH-5232 Villigen PSI, Switzerland}

\author{D. Sheptyakov}
\affiliation{Laboratory for Neutron Scattering ETHZ \& Paul-Scherrer
Institut CH-5232 Villigen PSI, Switzerland}

\author{S.G. Lushnikov}
\affiliation{Ioffe Physical Technical Institute, 26 Politekhnicheskaya,
194021, St. Petersburg, Russia}

\author{T.A. Shaplygina}
\affiliation{Ioffe Physical Technical Institute, 26 Politekhnicheskaya,
194021, 
St. Petersburg, Russia}

\date{\today}

\begin{abstract}
Neutron scattering investigations were carried out in PbMg$_{1/3}$Ta$_{2/3}$O$_3$ 
and BaMg$_{1/3}$Ta$_{2/3}$O$_3$ complex perovskites. The crystal structure   
of both compounds does not show any phase transition in the temperature range 
1.5~--~730~K. Whereas the temperature dependence of the lattice 
parameter of BaMg$_{1/3}$Ta$_{2/3}$O$_3$ follows the classical expectations, 
the lattice parameter of relaxor ferroelectric PbMg$_{1/3}$Ta$_{2/3}$O$_3$ 
exhibits anomalies. One of these anomalies is observed in the same temperature 
range as the peak in the dielectric susceptibility. We find that in 
PbMg$_{1/3}$Ta$_{2/3}$O$_3$, lead ions are displaced from the ideal positions in 
the perovskite structure at all temperatures. Consequently short-range order is 
present. This induces strong diffuse scattering with an 
anisotropic shape in wavevector space. The temperature dependences 
of the diffuse neutron scattering intensity and of the 
amplitude of the lead displacements are similar.
\end{abstract}

\pacs{77.80.-e, 61.12.-q}

\maketitle

\section{Introduction}
\label{intro}
Since complex AB$'_{x}$B$''_{1-x}$O$_3$ perovskites have  
two different ions in the B-sublattice, their properties 
are richer than the ones found in the classic ABO$_3$ 
counterparts~\cite{smolenski}. In particular, there is 
a subgroup of complex perovskites, relaxor ferroelectrics, 
which exhibit an anomaly in the dielectric susceptibility, 
broad in temperature and also frequency dependent. 
Unlike in usual ferroelectrics this anomaly does not link to 
macroscopic structural changes~\cite{smolenski,cross}. 
Despite numerous studies, a coherent understanding of 
relaxor ferroelectrics is still 
lacking~\cite{smolenski,cross,kleemann,viehland,blinc}. 
 
PbMg$_{1/3}$Nb$_{2/3}$O$_3$ (PMN) crystal is a model system 
to study relaxor behaviour~\cite{smolenski,cross,siny1}. 
The real part of the dielectric permeability $\varepsilon'$ 
of PMN has a maximum at T~$\sim$~270~K and frequency 
$\nu = 10$~kHz~\cite{smolenski}. The average crystal structure of 
PMN is cubic down to 5~K~\cite{husson1}. However, an 
applied electric field induces a structural phase transition 
at T$\sim210$~K~\cite{krainik}. PbMg$_{1/3}$Ta$_{2/3}$O$_3$ (PMT) 
is a well-known ferroelectric and it was among the 
first studied relaxors~\cite{smolenski}. The average 
structure is cubic in the range of examined  
temperatures and external electric fields~\cite{smolenski,lu}. 
In PMT, the real part of the dielectric susceptibility 
at $\nu$~=~10~kHz has a maximum at T$\sim170$~K, {\it i.e.}  
considerably lower than in other important relaxor crystals 
like PMN and PbZn$_{1/3}$Nb$_{2/3}$O$_3$ (PZN)~\cite{smolenski}. 
This suggests that temperature-induced anharmonicity    
less affects the properties of PMT in the range of the  
dielectric anomaly. Consequently, PMT can be 
considered as the simplest relaxor system known. Thus, 
understanding the physical mechanisms underlying the relaxor 
properties in a PMT crystal should be easier than in 
the related compounds. Unfortunately, PMT was studied much 
less then PMN and PZN. It is only 
recently that studies of PMT by X-ray diffraction~\cite{lu,akbas}, 
neutron scattering~\cite{seva1,seva2,seva3,seva4}, 
light scattering~\cite{seva2,serg1} and calorimetry~\cite{moriya} 
were carried out. On the other hand, low-dielectric permeability 
is found in many Ba and Sr-based complex perovskites~\cite{galasso}.
BaMg$_{1/3}$Ta$_{2/3}$O$_3$ (BMT) is an example of such a 
material~\cite{sinyBMT}. In spite of chemical 
disorder in the B-sublattice, BMT does not exhibit any lattice 
instability or relaxor-like dielectric 
anomaly~\cite{seva1,seva2,serg1,guo}. 
Note also that exchanging lead in PMT with barium in BMT suppresses  
the relaxor properties of the former crystal. Thus, comparing the 
properties of PMT and BMT might be extremely instructive. 

In this paper, we present results of neutron scattering studies in  
PMT and BMT in the temperature range 1.5~K$<$T$<$~730~K, that 
covers the dielectric maximum~\cite{smolenski} 
as well as the anomalies observed in the velocity and in the damping 
of the longitudinal acoustic phonon~\cite{serg1} in PMT. 
In Section~\ref{experiment}, we describe the experimental procedure. 
Section~\ref{results} starts with the explanation of the different 
models used for the crystal structure refinement of PMT and BMT and contains 
a discussion of the diffuse scattering measurements in PMT. 
Further, in $\S$~\ref{tempa}, we present the temperature dependences 
of the lattice parameters of both BMT and PMT. In $\S$~\ref{tempds},  
we describe the evolution of the temperature factors  
and show that there is a pronounced correlation 
between the diffuse scattering intensity and the Pb displacements 
in PMT. In $\S$~\ref{origin} we propose a possible approach to describe 
DS in relaxors.  Finally, in $\S$~\ref{alattice} we compare the 
temperature dependences of the lattice parameter of PMT and PMN. 
\section{Experimental}
\label{experiment}
The synthesis of the powder samples was described previously~\cite{seva1}. 
The mass of each sample was about 15 grams. The single 
crystal of PMT used in the present study was grown by 
the Czochralski method and has a size 
6.5 $\times$ 4.5 $\times$ 1.6 mm$^3$. The same crystal was used 
in previous dielectric, light and neutron scattering studies~\cite{seva3,serg1}. 

The measurements were performed at the neutron spallation 
source SINQ (PSI, Switzerland). The powder diffraction data were 
taken the the multi-detector high-resolution powder 
diffractometer HRPT~\cite{hrpt}. The polycrystalline samples 
of PMT and BMT were placed into cylindrical containers 
with 10~mm diameter and 50~mm height. Measurements were 
performed with the neutron wavelength $\lambda=1.1545 \rm~\AA$. 
This set-up gives access to more then 40 unique Bragg 
reflections in the range of scattering angles 
$\rm7^o$~--~164.5$\rm^o$. A typical exposure time was 
$\sim20$ minutes, but at few temperatures the exposure time 
was increased to $\sim4$ hours. 
The program Fullprof~\cite{rodriges} was used for crystal structure 
refinement using the Rietveld method. 
The neutron scattering measurements on a single crystal of PMT 
were performed at the cold neutron three-axis spectrometer 
TASP~\cite{tasp}. 
The (002) reflection 
of pyrolytic graphite (PG) was used to monochromate and 
analyze the incident and scattered neutron beams. 
The PG monochromator was vertically curved and a 
PG filter placed in the incoming neutron beam was used to suppress  
contamination from higher-order neutron wavelengths. 
The spectrometer was operated in the elastic mode with 
the neutron wavevector ${\bf k}_i=1.97 \rm~\AA^{-1}$. The horizontal 
collimation was guide$-80'-80'-80'$. The sample was mounted 
on an aluminum holder with the (h~h~0) and (0~0~l) 
Bragg reflections in the scattering plane. 
\section{Results}
\label{results}
\subsection{Structure Refinement}
\label{refinement} 

\begin{figure}[h]
 \includegraphics[width=0.5\textwidth]{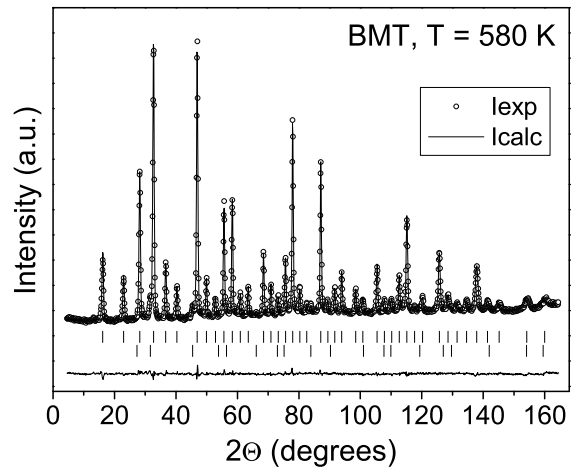}
 \caption{Neutron powder diffraction pattern of BMT collected at T~=~580~K. 
          Observed data points, calculated profile and difference curve are shown. 
          The two rows of ticks correspond to the calculated positions of diffraction
          peaks for BMT (upper) and MgO impurity (3\%).}          
\label{fig1}
\end{figure}

Figure~\ref{fig1} shows the observed and calculated diffraction 
patterns obtained for BMT  at T=580~K. For the refinements, 
the chemical composition of BaMg$_{1/3}$Ta$_{2/3}$O$_3$ was fixed 
at the stoichiometric values. Random occupation of the Mg/Ta ions 
over the B-sites of BMT perovskite structure was assumed. In a first approach, 
we used anisotropic temperature factors for oxygen. It turned out, 
however, that within the statistical errors, the ratio 
$\beta_{11}/\beta_{33}\sim1$ was constant even at high 
temperatures. Hence, for the subsequent refinements, the 
temperature factors were supposed to be isotropic. The 
crystal structure was found to be consistent with 
Pm$\bar{3}$m space group, where barium is located at 
(0~0~0), oxygen is at (0.5 0.5 0), and Mg/Ta ions are 
statistically distributed over the (0.5 0.5 0.5) position. 
At T~=~580~K the refinement procedure yielded a lattice constant 
$a$~=~4.10289(8)$\rm\AA$ and temperature factors 
B$_{iso} \rm (Ba)= 0.86(2) \AA^2$,  
B$_{iso} \rm (O)= 1.09(1) \AA^2$,  
B$_{iso} \rm (Mg/Ta)= 0.60(1) \AA^2$. 
This model gives a precise description of the experimental 
data ($\chi^2=2.52, \rm R_{p} = 3.49, R_{wp} = 4.44, R_{Bragg} = 3.57$) 
and was used to refine neutron diffraction patterns at other  
temperatures. We note that, despite a general tendency toward 
hexagonal ordering in Ba-containing complex 
perovskites~\cite{galasso,sinyBMT}, the synthesis method described in 
Ref.~\cite{seva1} produces cubic perovskites with ionic disorder in the B-site.

\begin{figure}[h]
  \includegraphics[width=0.5\textwidth]{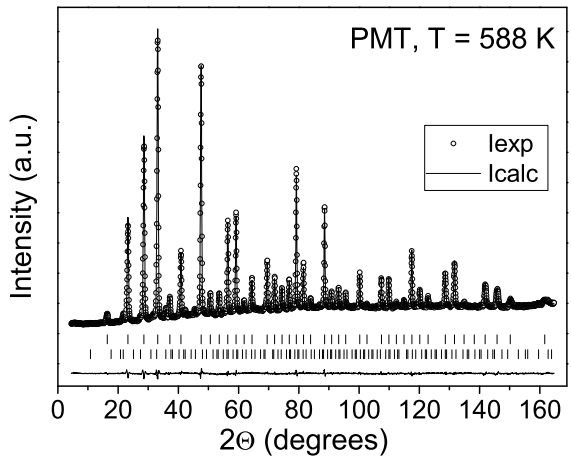}
  \caption{Neutron powder diffraction pattern of PMT collected at T~=~588~K. 
           Observed data points, calculated profile and difference curve are shown. 
           The two rows of ticks correspond to the calculated positions of diffraction
            peaks for perovskite PMT (upper) and pyrochlore impurity (5 \%).}
\label{fig2}
\end{figure}

Figure~\ref{fig2} shows the diffraction pattern obtained for PMT 
at T=588~K and treated with the Rietveld method. 
Different hypotheses for the structural model of PMT were tested 
during the refinement procedure. As in BMT,  
the cubic perovskite structure (space group Pm$\bar{3}$m) 
was assumed as a starting model. In this model, the refined values of the 
isotropic thermal parameters for lead and oxygen were 
found to be very high with B$_{iso} \rm (Pb)= 4.08(3) \AA^2$ and   
B$_{iso} \rm (O)= 1.73(1) \AA^2$, respectively. 
These values are much higher then those found for  
B$_{iso} \rm (Mg/Ta)= 0.58(7) \AA^2$, and for the isotropic 
temperature factors in the reference compound BMT. In particular, 
the large value of $\rm B_{iso}(Pb)$ is anomalous and indicates 
a possible structural disorder. A similar situation was  
observed before in many lead-based complex perovskites like 
PMN~\cite{husson2}, PMN+ 10\%PT (PMN-PT)~\cite{sb1}, 
PbSc$_{1/2}$Nb$_{1/2}$O$_3$~\cite{gvido1}, and also in 
PbFe$_{1/2}$Nb$_{1/2}$O$_3$~\cite{natasha} (the list is far from 
 complete). In all these examples, Pb was never found to be 
located exactly at the 1$a$ (0~0~0) Wyckoff position. Thus, to 
improve our structural model, both displacements of Pb from the 
1$a$ (0~0~0) special Wyckoff position and anisotropic temperature 
factors for oxygen ions were introduced. We considered the shifts of 
Pb 
along the $<$001$>$, $<$110$>$, and $<$111$>$ directions. 

At 588~K the values for the Pb-shifts together with the  
reliability factors R$_{Bragg}$ are tabulated in Table~\ref{table1}.   
Reliability factors for Pb-shifts along the $<$001$>$ direction 
are significantly worse than the corresponding value for the   
$<$110$>$ and $<$111$>$ directions. For the latter two directions, 
the quality of fit is nearly the same at all temperatures.
Displacing Pb along $<$110$>$ always gives a slightly better reliability 
factor and minimizes the value of the isotropic temperature factor of  
Pb.  However, we do not put a strict physical sense 
in the choice of this particular direction for Pb-shifts. Namely, as was 
pointed out in Ref.~\cite{sb4}, a preferable direction of the Pb-shifts 
can be artificial because of insufficiently high values in  
$\rm{(sin~\theta)}/\lambda$ (note that we reached 
$\rm{(sin~\theta_{max})}/\lambda=0.86~\AA^{-1}$). To our opinion, 
the quality of refinement for the different displacement
directions depends on the associated  
degeneracies: six-fold for the  
$<$001$>$, eight-fold for $<$111$>$ and twelve-fold for $<$110$>$. The
latter direction of shifts gives the closest 
approximation of a sphere. 
Hence,    
we used the $<$110$>$ direction to refine the data taken at other
temperatures.

To improve the structural model further, we introduced 
shifts for Mg/Ta ions and non-stoichiometric values for Pb 
and O content. However, the fit procedure converged to the nominal 
stoichiometric ratio. The introduction of shifts for 
Mg/Ta, while keeping all other parameters fixed, did not allow 
us to improve the structural model noticeably. Therefore, in the 
subsequent refinements, these parameters were fixed at their 
nominal values. The structural parameters obtained from the powder refinement 
at T=588~K are given in Table~\ref{table2}. We observe that on
the contrary to BMT,   
the thermal ellipsoid for oxygen is anisotropic 
($\beta_{11}/\beta_{33}\sim3.44$). 
We note also that despite introduction of displacements, the 
isotropic temperature factor for Pb is still 
quite important (see Table~\ref{table1}), being 2.5 times 
larger than B$_{iso}$(Mg/Ta) and 2 times larger 
than B$_{iso}$(Ba) in BMT. 
%
\begin{table}
\caption{Comparison between results of neutron powder refinement for different 
modeled  
directions of Pb shift at T=588~K in PMT.}
\label{table1}       
\begin{tabular}{lllll}
\hline\hline\noalign{\smallskip}
Direction & Value ($\rm\AA)$ & U$_{iso}\rm(Pb)$ (\AA{$^2$}) & R$_{Bragg}$ & $\chi^2$ \\
\noalign{\smallskip}\hline\noalign{\smallskip}
$<$001$>$ & 0.27(0.05) & 1.98(7) & 3.76 & 3.63\\
$<$110$>$ & 0.20(0.03) & 1.51(6) &3.28 & 3.44\\
$<$111$>$ & 0.16(0.04) & 1.68(6) &3.35 & 3.47\\
\noalign{\smallskip}\hline\hline 
\end{tabular}
\end{table} 
%
\begin{table}
\caption{Structural parameters for PMT  
obtained from refinement of data at T~=~588~K. Space 
group: Pm$\bar{3}$m. 
}
\label{table2}       
\begin{tabular}{lll}
\hline\hline\noalign{\smallskip}
Parameters & {\ }{\ }{\ }{\ }{\ }{\ }{\ }{\ } & Values  \\
\noalign{\smallskip}\hline\noalign{\smallskip}
a (\AA) & {\ }{\ }{\ }{\ }{\ }{\ }{\ }{\ } & 4.05103(8) \\
Pb shift (\AA) & {\ }{\ }{\ }{\ }{\ }{\ }{\ }{\ } & 0.20(0.03)\\
Pb U$_{iso}$ (\AA{$^2$}) & {\ }{\ }{\ }{\ }{\ }{\ }{\ }{\ } & 1.51(6) \\
Mg/Ta U$_{iso}$ (\AA{$^2$}) & {\ }{\ }{\ }{\ }{\ }{\ }{\ }{\ } & 0.66(1) \\
O $\beta_{11}$(\AA) = $\beta_{22}$(\AA) & {\ }{\ }{\ }{\ }{\ }{\ }{\ }{\ } &0.0340(3) \\
O $\beta_{33}$(\AA) & {\ }{\ }{\ }{\ }{\ }{\ }{\ }{\ } & 0.0101(4) \\
$\chi^2$ & {\ }{\ }{\ }{\ }{\ }{\ }{\ }{\ } & 3.44 \\
R$_p$ & {\ }{\ }{\ }{\ }{\ }{\ }{\ }{\ } & 2.88 \\
R$_{wp}$ & {\ }{\ }{\ }{\ }{\ }{\ }{\ }{\ } & 3.74 \\
\noalign{\smallskip}\hline
\hline
\end{tabular} 
\end{table}
\subsection{Diffuse scattering}
\label{diff} 
\begin{figure}[h]
  \includegraphics[width=0.5\textwidth]{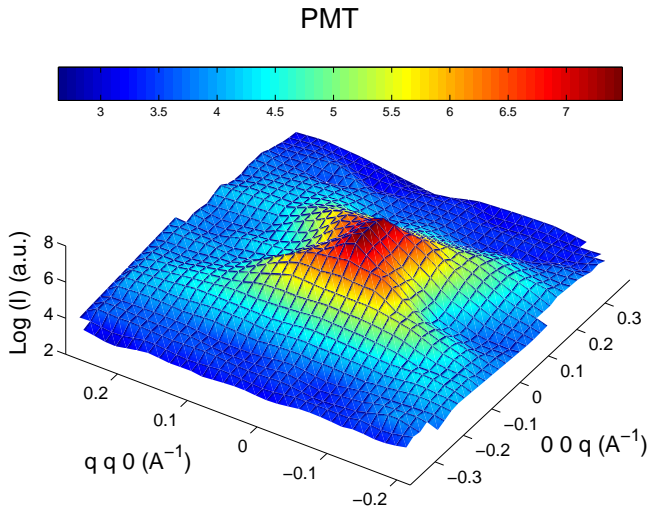}
  \caption{Surface of neutron diffuse scattering intensity around the 
          (110) Bragg peak of PMT at T = 140 K. Note that the intensity is given in a 
           logarithmic scale.}
\label{fig3}
\end{figure}

In the following, we discuss the influence of the diffuse scattering (DS) 
present on the powder diffraction pattern of PMT. Already in the 
earliest studies of PMN by X-rays~\cite{husson2} and neutron~\cite{sb2}
diffraction, diffuse scattering was found, the intensity of which grows 
as the temperature decreases~\cite{sb2}. Moreover, at low temperatures,   
DS is so strong that it becomes observable even in powder diffraction
patterns~\cite{husson2}.
Later work on PMN single crystals showed 
that the distribution of diffuse scattering is anisotropic in 
reciprocal space~\cite{sb3,shirane2,kreisel,shirane1,you1}. 
We have investigated the distribution of DS intensity in a single crystal 
of PMT at T~=~140~K in the vicinity of the (1~1~0) and (0~0~1) Bragg 
positions. As shown in Fig.~\ref{fig3}, the DS measured 
in the vicinity of (1~1~0) Bragg peak of 
PMT has a "butterfly" shape in reciprocal space similar to the DS 
observed in single crystals of PMN~\cite{shirane2}. Note, that 
the intensity of DS is mainly concentrated along the equatorial 
section transverse to the $<$110$>$ direction. Measurements in 
PMT around the (0~0~1) Bragg peak give a very similar shape of DS. In that
case, the most intense part of diffuse scattering is 
concentrated along the equatorial section transverse to the $<$001$>$ 
direction. It is obvious that the q- dependence of DS in PMT 
is complicated and to incorporate properly the DS into the structural 
model for powder refinement is an  extremely difficult problem. 
Hence, the DS occurring in the powder diffraction pattern was treated as background. 
\subsection{Temperature dependence of the lattice parameters of PMT and BMT}
\label{tempa}
\begin{figure}[h]
  \includegraphics[width=0.5\textwidth]{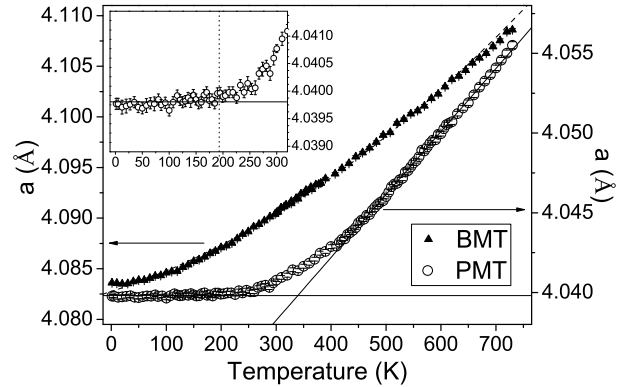}
  \caption{Temperature dependence of the lattice parameters of PMT and BMT. 
           The solid lines show the results of linear extrapolation for the a(T) 
           dependence of PMT. The dashed line shows the polynomial fit of a(T) 
           for BMT. The insert shows the dependence of the lattice parameter of 
           PMT in the region of temperatures around T$\sim$180 K. Invar-like 
           behaviour of PMT for $\rm T~<~180~K$ is obvious.}
\label{fig4}
\end{figure}  
Figure~\ref{fig4} shows that the temperatures dependences of the lattice 
parameters for BMT and PMT are qualitatively different. 
Starting from high temperature, the lattice parameter of BMT decreases almost  
linearly as the temperature decreases down to $\rm T\sim200~K$. 
There is a cross-over regime between this temperature and $\rm T\sim30~K$, 
below which the lattice parameter of BMT is approximately constant. 
The temperature dependence of the lattice parameter of BMT can be described 
by a quadratic polynom $\rm a_0+a_{01}{\times}T+a_{02}\times{T^2}$, where 
$\rm a_0=4.08275(8)\AA$, $\rm a_{01}=1.77(1)\cdot10^{-5}\AA\cdot{K^{-1}}$
and $\rm a_{02}=2.6(1)\cdot10^{-8}\AA\cdot{K^{-2}}$.   
Similar temperature dependences are commonly observed in solids
(see for instance~\cite{singh,geo} and references therein) and 
do not contain any signature of anomalies. This simply reflects 
that in a broad temperature range (1.5~K~--~730~K), the lattice of BMT
continuously expands. Hence, it appears that the presence of 
chemical disorder on the B-sublattice does not affect significantly 
the thermal expansion of the BMT lattice. 

The variation of the lattice parameters of PMT can be 
separated into three distinct temperature regimes. As in BMT, the 
lattice parameter decreases linearly from high temperatures 
($\rm da(T)/dT=3.8923\cdot10^{-5}\AA\cdot{K^{-1}}
T$) with a cross-over regime between T=420~K and T$\sim$180~K. 
Below this temperature, the lattice does not contract any more 
($\rm da(T)/dT=8.8\cdot10^{-7}\AA\cdot{K^{-1}}$). On the contrary 
to what is observed for BMT, a quadratic polynom does 
not reproduce the data shown in Fig.~\ref{fig4} and
it is necessary to include powers up to 6 to describe the temperature
dependence of the lattice parameter of PMT. 
It is clear that such a behaviour  
is anomalous. We remind the reader that the peak in the dielectric 
susceptibility of PMT appears around T$\sim$180~K~\cite{smolenski,lu}. 
Thus, the dielectric anomaly and 
the invar-like property in PMT are observed at the same temperature (see Insert of 
Fig.~\ref{fig4}). 

\section{Discussion}
\label{discussion}
\subsection{Correlation between the diffuse scattering intensity and 
Pb displacements in PMT} 
\label{tempds}
\begin{figure}[h]
  \includegraphics[width=0.5\textwidth]{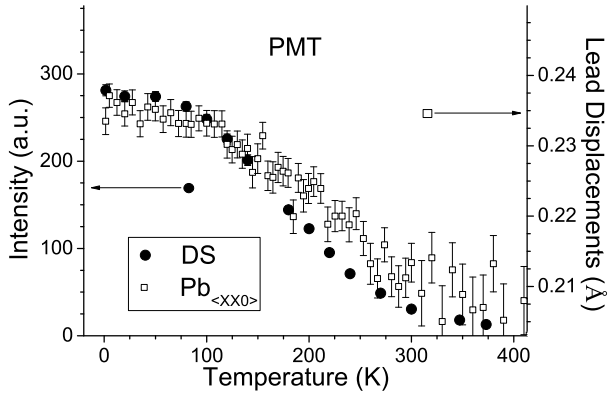}
  \caption{Temperature dependences of the integrated intensity of diffuse scattering 
          (DS) (black symbols) and amplitude of the Pb displacements (Pb$_{<XX0>}$) 
           (open symbols) in PMT.}
\label{fig5}
\end{figure}

We turn now to the description of the temperature dependence of the 
amplitude of Pb displacements (Pb$\rm_{<XX0>}$) and of the DS 
intensity in PMT shown in Fig.~\ref{fig5}. We measured the temperature dependence of the 
DS in PMT single crystal along [1,1,q]. In that direction,  
we find that the line-shape can be reproduced using a Lorentzian function~\cite{seva3}: 
\begin{equation}
\label{lor}
S(q)={I_0\over{2\pi}}\frac{\Gamma}{(q-q_0)^2+\Gamma^2},
\end{equation}
\noindent where $I_0$ is the integrated intensity, 
$\Gamma$ is the full width at half 
maximum (FWHM), and $q_0$ determines 
the center of the Lorentzian. Note that this 
approximation was used in Ref.~\cite{shirane1} to analyze the 
intensity of DS in PMN. We monitored the evolution of 
DS in PMT in the temperature range 1.5~K~--~375~K. At higher 
temperatures, the DS is extremely weak in intensity and broad 
in q-space, and the results obtained above $\rm T \sim 400 K$ 
cannot be analyzed reliably.  

Figure~\ref{fig5} shows the temperature behaviour of the integrated 
intensity of DS and of the refined values of the Pb$\rm_{<XX0>}$ shifts in the 
temperature range 
$\rm 1.5-412 K$. Note that the values of the Pb$\rm_{<XX0>}$ 
were derived from intensities of the Bragg peaks in the powder 
diffraction patterns whereas DS was extracted from the broad 
intensity distribution around these Bragg peaks {\it i.e.} from 
short-range correlations. The correlation length extracted from 
the FWHM of the DS line shape is found to increase continuously 
with decreasing temperature and was published 
previously~\cite{seva3}. There is a pronounced 
similarity between the increase of DS intensity and the 
displacement amplitude of Pb$\rm_{<XX0>}$ as the 
temperature decreases. In the temperature range $\rm1.5~K \sim 90~K$ 
both quantities are constant within statistical errors and decrease 
monotonously as the temperature is further increased. Finally, 
they tend to saturate above $\rm T~\sim~300~K$ although a 
very slow decrease of Pb$\rm_{<XX0>}$ above 412~K is not shown 
in Fig.~\ref{fig5} for better comparison with the DS intensity. 
The results shown in Fig.~\ref{fig5} strongly  suggest  
that lead displacements may be responsible for the occurrence 
of DS in PMT. There is however no linear relationship between 
the temperature dependence of  Pb$\rm_{<XX0>}$ and DS: 
while the DS intensity decreases by a factor of $\sim 30$ 
when the temperature increases from 1.5~K to 375~K, the change 
in Pb$\rm_{<XX0>}$ is of the order of $\sim$20\% only. At higher 
temperatures, lead displacements are still 
important, the DS is broad in q-space, 
since the correlation length is short~\cite{seva3}. In other words, while Pb 
displacements are {\it nearly uncorrelated} or {\it uncorrelated} 
the DS is weak (or totally absent). With decreasing 
temperature the Pb displacements increase and become {\it correlated}, 
which causes the appearance of the DS. Note also, that one 
cannot claim that only Pb displacements are responsible for 
DS. However, it was possible to satisfactorily fit the diffraction 
patterns of PMT assuming that only lead ions are displaced. 
This suggests that the main contribution to the DS  
in relaxor crystals originates from Pb ions. The recent 
observation of a temperature-dependent relaxation mode, coexisting 
with the central peak in PMN crystal~\cite{seva5}, also 
supports this hypothesis.  
\subsection{Possible description of the DS}
\label{origin}
We discuss now the resemblance observed for the temperature 
dependence of DS intensity and for Pb$\rm_{<XX0>}$. It seems 
that the q-dependence of DS in all relaxors, which have been 
studied so far, is very similar. For example, DS with a 
"Butterfly-like" shape is observed in PMN~\cite{shirane1,you1} and in 
PZN+8\%PbTiO$_3$~\cite{hlinka}. Furthermore, the intensity of DS 
is found to increase with decreasing temperature in 
PMN~\cite{sb1,sb2,shirane1} and PZN~\cite{toulouse,shirane3}, 
as here for PMT. From neutron and X-ray diffraction data,  
it is known that there is a general tendency that Pb 
displacements increase as the temperature decreases in complex 
perovskites (see {\it e.g.}~\cite{husson1,gvido1,sb4}). 
One can expect 
that the observed correlation between the DS intensity 
and the amplitudes of Pb displacements in PMT may help to 
create a model which describes short-range order  
in relaxors. In that vein, the calculations of Kassan-Ogly and Naish 
predicting an increase 
of diffuse scattering with decreasing temperature due to change in
ionic-oscillation amplitudes may be 
considered as a useful starting point~\cite{naish}.
\subsection{Lattice expansion}
\label{alattice}
High-resolution neutron diffraction allowed us to study the 
temperature dependence of the lattice constant $a$ in PMT from  
41 unique Bragg reflections in the temperature range 
$\rm1.5~K-730~K$ with temperature step of $\rm\sim10~K$. 
We found changes of slope in $a(T)$ at $\rm\sim420~K$ and 
$\rm\sim180~K$ which do not strictly agree with previous 
temperature studies of the lattice parameters in PMT. 
In Ref.~\cite{lu}, an anomaly was observed first at 
T=573~K and the lattice parameters were found to remain 
constant below T=280~K. We ascribe the discrepancy between 
our data and those by Lu {\it et al.}~\cite{lu} to the facts that, 
first, the latter measurements were carried out using 
three Bragg peaks of low Miller indices and, second, the 
number of temperature points was much reduced as compared 
to the present study. Thus, to our opinion there is no 
anomaly in the lattice parameter of PMT which 
could be directly connected with the Burns temperature 
$\rm T_d\sim570~K$~\cite{comment}. Interestingly, the 
change of slope of $a(T)$ in PMT at high temperatures 
occurs at temperatures where we are able first to observe 
the diffuse scattering. In other words, deviation from linear 
law in $a(T)$ of PMT crystal appears when the shifts of Pb 
ions become correlated. Note, that the so-called "excess"
specific heat in PMT also appears at 
$\rm T\sim420~K$~\cite{moriya}.   

It is useful to compare the behaviour of the lattice 
parameters of PMT and PMN~\cite{husson1,sb1,darling} in the 
temperature ranges where dielectric anomalies are observed. 
In the first neutron measurements by 
Husson {\it et al.}~\cite{husson1}, the lattice parameter of 
PMN was observed to remain roughly constant below  
$\rm T\sim300~K$. However, recent studies have 
revealed a more complicated behaviour of $a(T)$ in PMN. In 
Refs.~\cite{sb1,darling} a change of slope in $a(T)$ 
was found at $\rm T\sim~350~K$ and at $\rm T\sim~200~K$. In 
the temperature range 200~--~350~K, Dkhil 
{\it et al.}~\cite{sb1} reported a freezing of the 
lattice parameters, similar to what we found in PMT below 
$\rm\sim 180~K$. An important qualitative difference in the 
behaviors of the lattice parameters of PMN and PMT is that 
below $\rm T\sim200~K$, $a(T)$ for PMN decreases again as the 
temperature is lowered, whereas for PMT such a change of 
regime is absent. We remind that at T$\sim210$~K PMN has 
an additional phase transition in an applied electric 
field~\cite{krainik}. However, even without application 
of an external electric field, properties like damping of 
longitudinal acoustic phonons~\cite{serg3}, and the width of 
quasielastic component in light scattering spectra~\cite{serg4} 
exhibit anomalies in PMN in the vicinity of this temperature. 
Thus, it is possible that this lattice instability 
is linked with the additional change of $a(T)$ of PMN as the 
temperature decreases below 200~K.

In the following we discuss possible causes for 
the anomalous temperature dependence of the lattice parameter in PMT.
As it is well-known (see for instance~\cite{baron}), 
the unit cell volume $V$ depends on temperature $T$ as  
\begin{equation}
\label{termodynamics}
\frac{dV}{V}=\frac{\kappa_TC_V}{V}\gamma(T,V)dT, 
\end{equation}
\noindent where $\kappa_T$ is the isothermal compressibility, 
$\gamma(T,V)$ the Gr\"uneisen function, and $C_V$ the 
heat capacity taken at constant volume. 
However, it is not easy to understand the low-temperature 
saturation of $a(T)$ below the temperature of the maximum 
in $\varepsilon'$. The problem is related to the large 
number of observed anomalies in the vibration spectra of PMT 
(as well as in other relaxors) for temperatures close to the 
dielectric maximum, like 
\begin{itemize} 
\item increase in the generalized density of states at 
low energies~\cite{seva1,seva2}, 
\item anomalies in the behaviour of the longitudinal acoustic 
phonons~\cite{serg1,serg2},
\item excess contribution to the specific heat~\cite{moriya,seva6}. 
\end{itemize}
Hence, to estimate the influence of each of this anomalies on 
the behaviour of $a(T)$ using {\it e.g.} Eq.~(\ref{termodynamics}) 
is an extremely complicated problem. We note only that an  
invar-like behaviour generally can be described e.g 
by compensation of anharmonic behaviour of 
different phonon modes~\cite{grimval} or striction. 
The electrostriction mechanism originally proposed in Ref.~\cite{cross} 
can also give some contribution in the behaviour of $a(T)$. Since $a(T)$ is 
linear totemperatures well below $T_{d}$, 
it is unlikely that electrostriction is the only mechanism 
responsible for the temperature evolution of the lattice parameters in PMT.  

\section{Conclusion}
To summarize, an extensive neutron study has been carried out in 
PMT and BMT complex perovskites in the temperature range 1.5~K -- 730~K. 
The surface of diffuse scattering in a PMT single-crystal has been measured 
at T~=~140~K. Crystal structure parameters have been refined for both compounds.
We found that lead ions are shifted from the 1$a$ (0~0~0) Wyckoff position 
of the ideal perovskite structure and that the thermal motion parameter 
of oxygen is anisotropic in PMT. The lattice parameter of BMT does not 
exhibit any anomalous behaviour. On the other hand, the lattice parameter 
of PMT shows two anomalies. One of them appears around T~=420~K 
which corresponds to the appearance of diffuse scattering and of the 
excess specific heat in PMT. The second anomaly is found close to the 
maximum of the dielectric susceptibility at T=180~K. An obvious correlation 
in the behaviour of the lead displacements and of the neutron 
diffuse scattering intensity is observed in PMT. This indicates that 
lead displacements play a major r\^ole in the occurrence of 
diffuse scattering in PMT and probably also in other relaxor ferroelectrics. 

\begin{acknowledgments} 
The authors would like to thank Dr. P. Fischer for useful 
discussions and N.V. Zaitzeva for X-rays characterization of powder 
samples. This work was performed at the spallation neutron source 
SINQ, Paul Scherrer Institut, Villigen (Switzerland) and was 
partially supported by RFBR Grant No. 02-02-17678 and by Grant of 
President RF ss-1415.2003.2. 
\end{acknowledgments}

%

\end{document}